\documentclass [article]{nature}
\usepackage{float}
\usepackage{units}
\usepackage{graphicx}
\usepackage{color}
\usepackage{epstopdf}
\usepackage{amsmath}
\usepackage{amsfonts}
\usepackage{amssymb}
\usepackage{dcolumn}
\usepackage{bm}
\usepackage{latexsym}

\title{Field-induced resistivity plateau and
unsaturated negative magnetoresistance in topological semimetal
TaSb$_2$}

\author{Yuke Li$^{1}$\footnote[1]{yklee@hznu.edu.cn},
Lin Li$^{1}$, Jialu Wang$^{1}$, Tingting Wang$^{1}$,
Xiaofeng Xu$^{1}$, Chuanying Xi$^{2}$, Chao Cao$^{1}$, Jianhui Dai$^{1}$\footnote[2]{daijh@hznu.edu.cn}}

\date{\today}
\begin{document}
\maketitle

\begin{affiliations}
 \item Department of Physics and Hangzhou Key Laboratory
of Quantum Matter, Hangzhou Normal University, Hangzhou 310036,
China
 \item High Magnetic Field Laboratory, Chinese Academy of Sciences, Hefei, Anhui
230031, China
\end{affiliations}

\begin{abstract}
\indent{}

Several prominent transport properties have been identified as key
signatures of topological materials\cite{TI,Qi}. One is the
resistivity plateau at low temperatures as observed in several
topological insulators (TIs)\cite{Ren,Jiashuang,Wolgast13,Kim13};
another is the negative magnetoresistance (MR) when the applied
magnetic field is parallel to the current direction as observed in
several topological semimetals (TSMs) including Dirac semimetals
(DSMs)\cite{Kim2,ZrTe5,Xiongjun,CLi,CZhang,ShenSQ} and Weyl
semimetals (WSMs)\cite{TaP,GFChen,ShuangJia,Zwang,JHDu}. Usually,
these two exotic phenomena emerge in distinct materials with or
without time reversal symmetry (TRS), respectively. Here we report the discovery
of a new member in TSMs, TaSb$_2$, which clearly exhibits both of
these phenomena in a single material. This compound crystallizes in
a base-centered monoclinic, centrosymmetric structure, and is
metallic with a low carrier density in the zero field. While
applying magnetic field it exhibits insulating behavior before
appearance of a resistivity plateau below T$_c=$13 K. In the plateau
regime, the ultrahigh carrier mobility
and extreme magnetoresistance (XMR)
for the field perpendicular to the current are observed as in
DSMs\cite{NaBi1,NaBi2,NaBi3,CdAs2,CdAs3} and
WSMs\cite{TaAs,TaAs2,TaAs3,TaAs4,NbAs,NbAs2,NbP}, in addition to a
quantum oscillation behavior with non-trivial Berry phases. In
contrast to the most known DSMs and WSMs, the negative MR in
TaSb$_2$ does not saturate up to 9 T, which, together with the
almost linear Hall resistivity, manifests itself an electron-hole
non-compensated TMS. These findings indicate that the resistivity
plateau could be a generic feature of topology-protected metallic
states even in the absence of TRS and compatible with the negative
MR depending on the field direction. Our experiment extends a
materials basis represented by TaSb$_2$ as a new platform for future
theoretical investigations and device applications of topological
materials.

\end{abstract}

\indent{}

Diverging longitudinal resistivity with decreasing temperature is
the most apparent transport property of a simple band insulator in
distinction to any metals. However, such a distinct feature ceases
to stand in various topological insulators due to the existing
metallic surface states\cite{TI,Qi}. Nevertheless there are other
transport phenomena, such as resistivity plateau and negative MR,
which may distinguish at least ideal TIs and
 ideal TSMs from topologically trivial materials.
 In an ideal 3D TI where the bulk states are
completely gapped out near the Fermi level, a resistivity plateau
can be clearly established because the only participating surface
states are TRS protected and thus robust to disorders, leading to a
saturation of resistivity in the low temperature
regime\cite{Ren,Jiashuang,Wolgast13,Kim13}. In TSMs without
coexisting bulk Fermi surfaces such as an ideal WSM, on the other
hand, the bulk excitations come from the two separated Weyl nodes in
momentum space which are chiral in nature owing to the lack of TRS
or inversion symmetry\cite{WanX,Xu,Weng}. When the applied magnetic
field is parallel to the electric field direction, the density of
the right/left chiral excitations increases/decreases accordingly as
a consequence of the chiral anomaly\cite{ABL,ABL2,Chiral}, resulting
in a non-dissipative current from the left to right nodes along the
field direction, hence an unconventional negative MR appears.

Because the resistivity plateau and negative MR are opposite
consequences for systems with or without TRS in the ideal situations
mentioned above, coexistence of the both in a single material is
unlikely or very difficult. Of course, these features should be more
intriguing but much involved in realistic topological materials,
and, TRS itself is not the only origin relevant to the resistivity
plateau on a general ground. In two-dimensional electron gases or
semiconducting films like graphite\cite{Abrahams,Kempa}, for
instance, the resistivity seemingly saturates after a field-induced
metal-insulator transition while a clear resistivity plateau at
lower temperatures was not reported. More recently, a field-induced
plateau has been observed in LaSb, a potentially new candidate of
TIs\cite{LaSb}. While the interpretation of all these and related
behaviors in topological materials remains a theoretical challenge,
materials realizations of these effects are highly desirable not
only in confronting this challenge but also in technique
applications.

\indent{}

Here we report the discovery of a new TSM TaSb$_2$,
which crystallizes in a monoclinic structure with centrosymmetric
space group $C_{12/m1}$ (SI-Fig. S1). The longitudinal resistivity
and Hall effect were measured for various magnetic fields applied
along different directions. The semiconducting behavior is
associated with a low carrier density and a field-induced
metal-to-insulator-like transition. High quality of the measured
samples is indicated by ultrahigh mobility and XMR. The topological
nature of the compound is evidenced by the Shubnikov de Haas (SdH)
oscillation measurement as well as band structure calculations
(SI-Fig. S6). Surprisingly, both the field-induced resistivity
plateau and unsaturated negative MR can be clearly observed in this
compound as illustrated in the following. Therefore, our experiment
shows that the monoclinic TaSb$_2$ represents a potentially new
class of topological materials exhibiting all these novel phenomena.

\indent{}

The magneto-transport properties of the sample 1 for TaSb$_2$ is
summarized in Figure 1, where the applied magnetic field ${\textbf
B}$ is parallel to the c-axis, and normal to the current. Figure 1a
describes the temperature dependent longitudinal resistivity $\rho
(T)$ at various fields ($\mu_0$H up to 9 T). At ${\textbf B} = 0$,
$\rho(T)$ exhibits highly metallic behavior down to 2 K. Its value at
300 K is 7.4 $\times10^{-2}$ m$\Omega$cm, about one order smaller
than that of WTe$_2$, an electron-hole compensated semimetal with XMR\cite{Cava}. For nonzero field B (= $\pm|{\textbf B}|$), $\rho(T)$
firstly decreases, then increases rapidly upon cooling, showing a
crossover to insulating behavior. The insulating behavior becomes
prominent when B $>$ 1 T.
Remarkably, $\rho (T)$ saturates at the low temperature regime,
developing a resistivity plateau which is clearly exhibited in
Figure 1b with a logarithm scale.


This finding reminds us of the compound SmB$_6$, an important candidate of TIs exhibiting a
very similar resistivity plateau at zero
field\cite{Wolgast13,Kim13}. In that compound, the topological
non-trivial surface states have been evidenced by both
experiments\cite{Kim,Li13} and band structure
calculations\cite{Lu13}, so the origin of the plateau is best
understood as due to the surface states. The similar plateau in
$\rho$(T) but under applied fields has been also observed in LaSb and
other semimetal compounds\cite{LaSb,NbSb2}, the former was
predicted as a new kind of TIs\cite{FuL}. The plateau of TaSb$_2$ onsets at $T
=$ 13 K, almost three times larger than $T =$ 5 K in SmB$_6$, and
comparable to $T =$ 15 K in LaSb.

Figure 1c shows temperature dependence of the derivative
$\partial{\rho}(T)/\partial{T}$ at different fields. A clear drop is
seen for B $\gtrsim$ 2 T and becomes prominent for larger B.
The temperature location of the drop peak, $T_i$, where
$\partial^2{\rho}(T)/\partial{T^2}=0$, is the inflection point where
crossover from insulating behavior to plateau takes place. The
metal-insulating-like transition takes place at an elevated
temperature, $T_m$, where $\partial{\rho}(T)/\partial{T}=0$. The
inset of the Figure 1c shows the evolution of $T_m$ and $T_i$ vs.
magnetic field B, indicating that $T_m$ increases monotonously and
$T_i$ remains almost unchanged. This silent feature implies that the
insulating behavior is of magnetic origin while the plateau may be
of topological one. If this is true, the later could be there at zero field. Indeed, we find that the lines of $T_m$ and
$T_i$ seem to merge at B $= 0$ T, suggesting a possible plateau
there. In order to estimate the insulating gap, $E_g$, Figure 1d
plots the $Log(\rho)$ as a function of $T^{-1}$ in the range of $T_i
< T < T_m$ using $\rho(T)= \rho_0 \exp(E_g/K_B T)$ with constant
$\rho_0$. The inset in Figure 1d shows the fitted activation energy
gap $E_g\propto \sqrt {B}$. This is consistent with the gap opening
in relativistic Dirac electrons by the magnetic field.

\indent{}

Figure 2 plots the field dependence of MR, where negative B means
the opposite direction. As shown in the upper panel of Figure 2a,
the XMR of TaSb$_2$ at low temperatures is exhibited when the field
is perpendicular to the current, reaching 15000\% at $T =$ 2 K and B
$=$ 9 T. This MR is quadratic for low field and almost linear for
larger B without saturation, similar to many known semimetallic
materials including TaAs(P), NbAs(P) and
WTe$_2$\cite{TaAs,TaP,NbAs,NbP,Cava}. Upon heating, the MR decreases
slowly below 10 K, but drops quickly at higher temperatures. The
inset in the upper panel of Figure 2a shows a window for 7 T $\leq$
B $\leq$ 9 T where the SdH oscillations are clearly exhibited at $T
=$ 2 K and 5 K, respectively.

We also measured the MR by applying the magnetic field parallel to
the current ${\textbf B}||{\textbf I}$, as shown in the
low panel of Figure 2a. It is remarkable that (i) the MR in the whole
temperature regime is less than 100\%, much smaller than that in the
case of $ {\textbf B}\bot {\textbf I}$ shown previously; (ii) in the
low temperature regime such as $T =$ 2 K and 10 K, the MR is positive
and parabolic for B $\lesssim $ 6.5 T;  (iii) while for B $\gtrsim $ 6.5
T, it becomes negative and decreases with increasing magnetic field
without saturation up to B $=$ 9 T; (iv) when $T \gtrsim $ 50 K, the
MR is very small and positive, increasing monotonously with field.

Figure 2b describes the field dependence of MR at a fixed
temperature $T =$ 2 K with different angles $\theta $ between
the magnetic field and current. The overall profile is nearly
symmetric under B $\rightarrow$ $-$ B. By rotating $\theta =
90^\circ\rightarrow 0^\circ$, the MR drops quickly, in particular
when approaching $\theta \approx 10^\circ $. The tendency of
negative MR for smaller $\theta$ is further illustrated in the low
panel of Figure 2b.  At B $=$ 9 T, the MR turns to negative for
$\theta \lesssim 4^\circ$. The window for the negative MR is limited
from this point to B $\gtrsim  $ 6.5 T at $\theta $ = 0$^\circ$. The
similar narrow $\theta$ window was observed in Na$_3$Bi and TaP
compounds\cite{TaP,Xiongjun}. The unsaturated negative MR, which
decreases monotonically with fields reaching about -74\% at B $=$ 9, is
a remarkable feature compared to the prototype DSMs and WSMs such as
Na$_3$Bi\cite{Xiongjun}, Cd$_3$As$_2$\cite{CLi,CZhang,ShenSQ}, TaAs\cite{GFChen}, NbP\cite{Zwang}, where the negative MR is limited
not only in a narrow window of $\theta$, but also in a window of B,
namely, the negative MR will saturate and return to positive for
larger $B$ even for $\theta = 0^{\circ}$. In view of chiral anomaly,
the MR should be always negative as long as ${\textbf B}||{\textbf
I}$ (or $\theta = 0^{\circ}$). If imperfect alignment of the
magnetic field and the current in samples can be fully excluded, the
limited window in B should be due to the
disorder-induced weak localization. So the unsaturated negative MR
observed here implies the consistency with the chiral
anomaly interpretation and the high quality of the measured
samples.

\indent{}

Figure 3 maps the Hall effect and the SdH oscillations. The magnetic field dependence of Hall resistivity at
various temperatures is displayed in the main panel of Figure 3a. The
negative slope for all of curves implies that electron-type carriers
dominate the transport from 300 K to 2 K. The $\rho_{xy}$ displays
an overall linear dependence, and only slightly deviates from linear
at around 9 T as $T \leq$ 20 K. The Hall coefficient $R_H$ versus
temperature at 6 T and 9 T is plotted in the inset of Figure 3a. $R_H$
is always negative and drops soon below 50 K without a sign change.
Notice that the strong nonlinear behavior in $\rho_{xy}$ and the
sign changed $R_H$ in several semimetals \cite{NbP,TaAs2,Cava} have
been regarded as indications of the electron-hole compensation.
Obviously, this interpretation cannot apply to the present TaSb$_2$
sample.

Accordingly, the calculated carrier concentration is $n_e\sim 3.2
\times$10$^{20}$ cm$^{-3}$ using $n_e = 1/eR_H(T)$ and the estimated
mobility $\mu_e \sim 1.96\times$ 10$^4$ cm$^2$V$^{-1}$S$^{-1}$ at 2
K. Hence,  TaSb$_2$ has a low carrier density of electron-type but
with a high carrier mobility, similar to the results of LaSb\cite{Cava}.  A direct consequence of these for the
SdH oscillation is shown in Figure 3b for $\rho_{xx}(B)$ at 1.8 K,
4.2 K and 8 K, respectively. We extract the SdH oscillations using
$\rho_H = \rho_0[1+ A(B, T)cos 2\pi(S_F/B-\gamma+\delta)]$, with
$\rho_0$ being the non-oscillatory part, A(B, T) the amplitude,
$\gamma$ the Onsager phase, $F = \frac{\hbar}{2e\pi}S_F$ the
frequency, and $S_F$ the cross-section area of the Fermi surface
associated with the Landau level index $\emph{n}$. The background is
subtracted using a polynomial fitting. The obtained
$\Delta\rho_{xx}$ as a function of $B^{-1}$ is then plotted in the
inset of Figure 3b.
Two oscillation sets can be extracted (SI-Fig. S5), corresponding to
two frequencies: a small oscillation frequency at $F_{\alpha} =$ 220
T, and a second frequency at $F_{\beta} =$ 465 T with its harmonic
$F_{2\beta} \approx$ 930 T and $F_{3\beta} \approx$ 1377 T, as shown
in Figure 3c. The Berry phase $\Phi_B$ can be identified via the
relation $\gamma = 1/2 - \Phi_B/2\pi$, so $\Phi_B$ is non-trivial
when $\gamma \neq $ 1/2. We thus plot the Landau fan diagram in
Figure 3d, and count down to $n = $8 and 16 for F$_\alpha$ and
F$_\beta$, respectively, up to magnetic field 30 T. Linear fitting
of n versus B$^{-1}$ yields the Onsager phase of $\gamma_\alpha =$
0.29 and $\gamma_\beta =$ 0.2, respectively, far away from one half, indicative of
non-trivial $\pi$ Berry phases in TaSb$_2$. The slopes of the linear
fitting are F$_{\alpha fit} =$ 228 T and F$_{\beta fit} =$ 461 T,
respectively, consistent with the experimental values. The
non-trivial Berry phases identified here indicate the topological
origin  of the resistivity plateau. This conclusion is also
consistent with the first principle calculations for the electronic
band structure of TaSb$_2$ as described in the Supplemental
Information. The calculations indicate that TaSb$_2$ has a small
bulk Fermi surface and a Dirac cone near the Fermi level, both of
them are contributed by the partially occupied, topologically
non-trivial electronic bands.

\indent{}

In summary, we report the discovery of a TMS TaSb$_2$ with a monoclinic crystal structure. It undergoes
a metal-insulator-like transition induced by magnetic field upon
cooling. Yet this compound shares a number of excellent transport
properties including the positive XMR and high mobility. In the
low temperature regime, it exhibits both the resistivity plateau and
unsaturated MR when the applied field is perpendicular and parallel
to the current, respectively. The topological property is manifested
by the non-trivial Berry phases in the SdH oscillations as well as
the band structure calculations.

Given the fact that the resistivity plateau and the negative MR are
characteristic features of ideal TIs and TSMs with and without TRS,
respectively, the coexistence of these two distinct phenomena in
TaSb$_2$ is a rather remarkable observation. It implies that the
resistivity plateau may be a generic feature of a wide class of
topological materials possessing metallic surface states even in the
absence of TRS. It also raises a challenge to understand the different fates of bulk Fermi surfaces, Dirac cone excitations, as well as metallic surface states in the presence of magnetic field.
 All these novel features, together with the
field-induced metal-insulator transition, the XMR and the high
carrier mobilities, suggest TaSb$_2$ as an interesting new platform
of topological materials for future theoretical investigations and
device applications.

\begin{methods}
\indent{} High quality single crystals of TaSb$_2$ were grown via
chemical vapor transport reaction using iodine as transport agent.
Ploycrystalline samples of TaSb$_2$ have been first synthesized by
solid state reaction using high purified Tantalum powders and
Antimony powders in a sealed quartz tube. The final powders were ground
thoroughly, and then were sealed in a quartz tube with a transport
agent iodine concentration of 10 mg$/cm^3$ . The single crystals
TaSb$_2$ were grown by a chemical vapor transport in a temperature
gradient of 120 $^{\circ}{\rm C}$ between 1120 $^{\circ}{\rm C}$ -
1000 $^{\circ}{\rm C}$ for 1-2 weeks. X-ray diffraction patterns
were obtained using a D/Max-rA diffractometer with CuK$_{\alpha}$
radiation and a graphite monochromator at the room temperature. The
single crystal X-ray diffraction determines the crystal grown
orientation. The composition of the crystals were obtained by energy
dispersive X-ray (EDX) spectroscopy. No iodine impurity can be
detected in these single crystals.

The (magneto)resistivity and Hall coefficient measurements were
performed with a standard four-terminal method covering temperature
range from 2 to 300 K in a commercial Quantum Design PPMS-9 system
with a torque insert. The deviation of MR and Hall measurements
associated with misalignment of the voltage leads could be corrected
by reversing the direction of the magnetic field. The high
magnetic field resistivity and Shubnikov-de Haas oscillations were
measured up to 31 T at Hefei High Magnetic Field Laboratory. To
check those experimental data, we performed resistivity and Hall
effect measurements in both sample 1 and sample 2 with different
RRR (See SI). The sample 2 with lower RRR also shows the same
experimental results (See SI).

\indent{}The electronic structure calculations were performed in the
framework of density functional theory using the Vienna Abinitio
Simulation Package (\textsc{VASP})\cite{vasp1,vasp2} with
projected augmented wave (PAW) approximation and Perdew Burke
Ernzerhoff (PBE) flavor of the generalized gradient approximation
(GGA). A 400 eV plane-wave energy cut-off and a $8\times 8\times 5$
$\Gamma$-centered K-grid was chosen to ensure the convergence to 1
meV/atom. The Fermi surface was then obtained by extrapolating the
DFT band structure to a dense K-grid of $100\times 100\times 100$.
The topological indices were calculated following the parity-check
method\cite{Fu} proposed by Fu {\it et al.}.

\end{methods}

\begin{addendum}

\item [Acknowledgements]
This research was supported in part by the NSF of China (under
Grants No. 11274006 and No. 11274084) and the National Basic
Research Program (under Grant No. 2014CB648400). A portion of this
work was performed on the Steady High Magnetic Field Facility, the
High Magnetic Field Laboratory, CAS. We would like to thank Jinke
Bao, Yanliang Hou, Yupeng Li, Hangdong Wang, and Zhen Wang for technique assistances. The authors are grateful to
Guanghan Cao, Xi Dai, Xiao-Yong Feng, and Mingliang Tian for stimulating discussions.

 \item[Author Contributions]
Y. Li designed the research. L. Li synthesized the samples. L. Li
and Y. Li performed most measurements. J. Wang and T. Wang performed part of measurements. C. Xi performed the SdH oscillations measurements
at high magnetic fields. C. Cao performed
 the band structure calculations. C. Cao, J. Dai, Y. Li, and X. Xu
discussed the data, interpreted the results, and wrote the paper.

 \item[Competing Interests]
The authors declare no competing financial interests.
 \item[Correspondence]
Correspondence and requests for materials should be addressed to
Yuke Li(email: yklee@hznu.edu.cn).

\end{addendum}

\clearpage

\begin{figure}
\includegraphics[angle=0,width=16cm,clip]{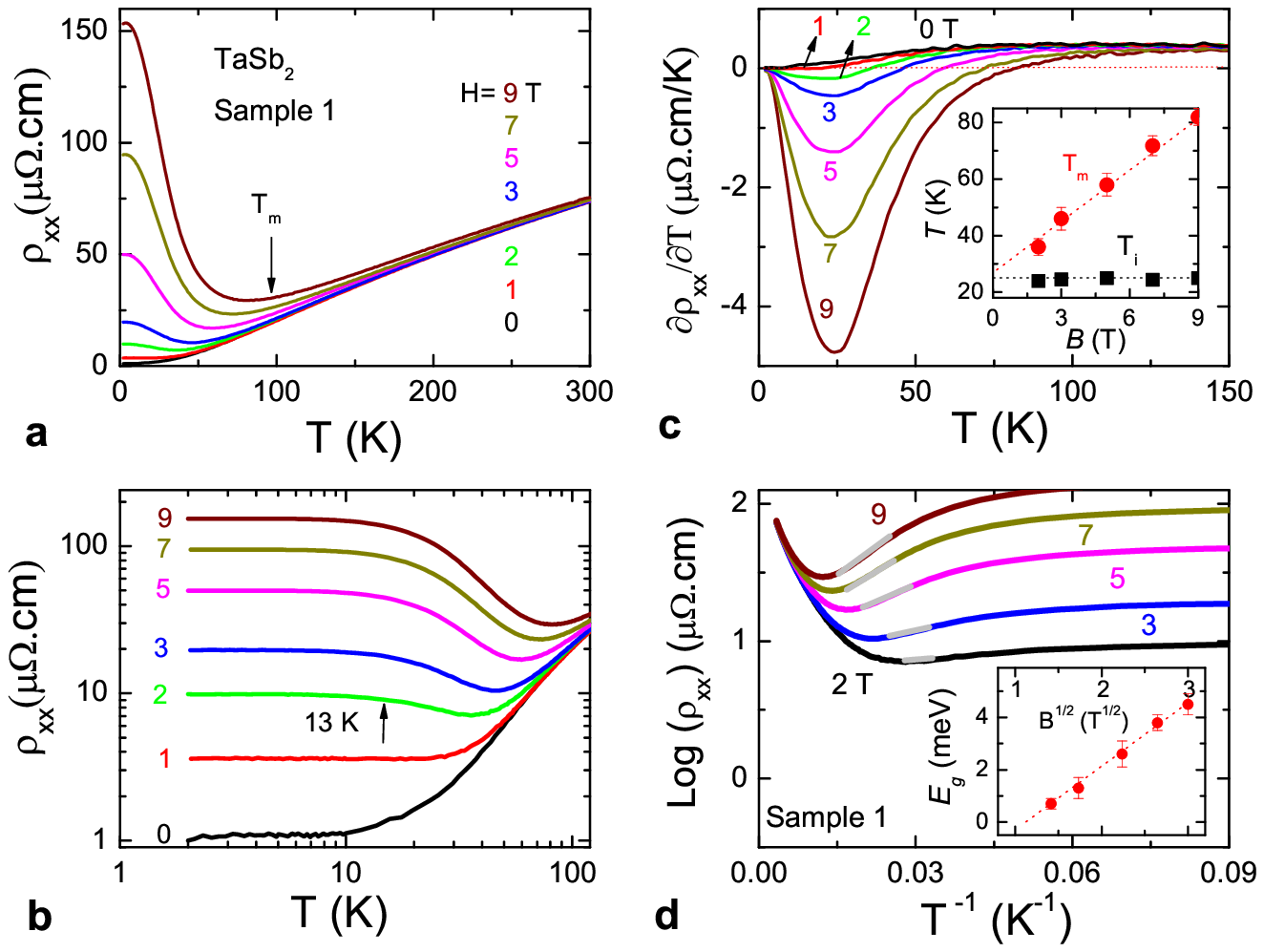}
\caption{{\bf Temperature dependence of resistivity in sample 1 for
TaSb$_2$.} $\textbf{a}$, Resistivity of TaSb$_2$ as a function of
temperature in several magnetic fields ($B = 1,2,3,5,7,9$ T.)
perpendicular to the current. The $\textbf{b}$ shows a clear plateau
resistivity at low temperature above 2 T. $\textbf{c}$, Temperature
dependence of $\partial{\rho}/\partial{T}$ at several
magnetic fields. The peak in $\partial{\rho}/\partial{T}$ is
defined as the inflection point $T_i$ at each field. The sign change
in $\partial{\rho}/\partial{T}$ symbolizes the resistivity
minimum at $T_m$. Inset shows $T_m$ and $T_i$ vs. magnetic fields. The dashed lines guide your eyes. $\textbf{d}$,
Log($\rho$) plotted as a function of the inverse temperature used to
extract the activation gap ($E_g$) in TaSb$_2$. The cyan lines show
the region of the linear fits. The inset shows field
dependence of the activation gap values ($E_g$). Dashed line shows
the linear behavior $E_g$ $\propto$ $B^{1/2}$.}
\end{figure}

\clearpage

\begin{figure}
\includegraphics[angle=0,width=16cm,clip]{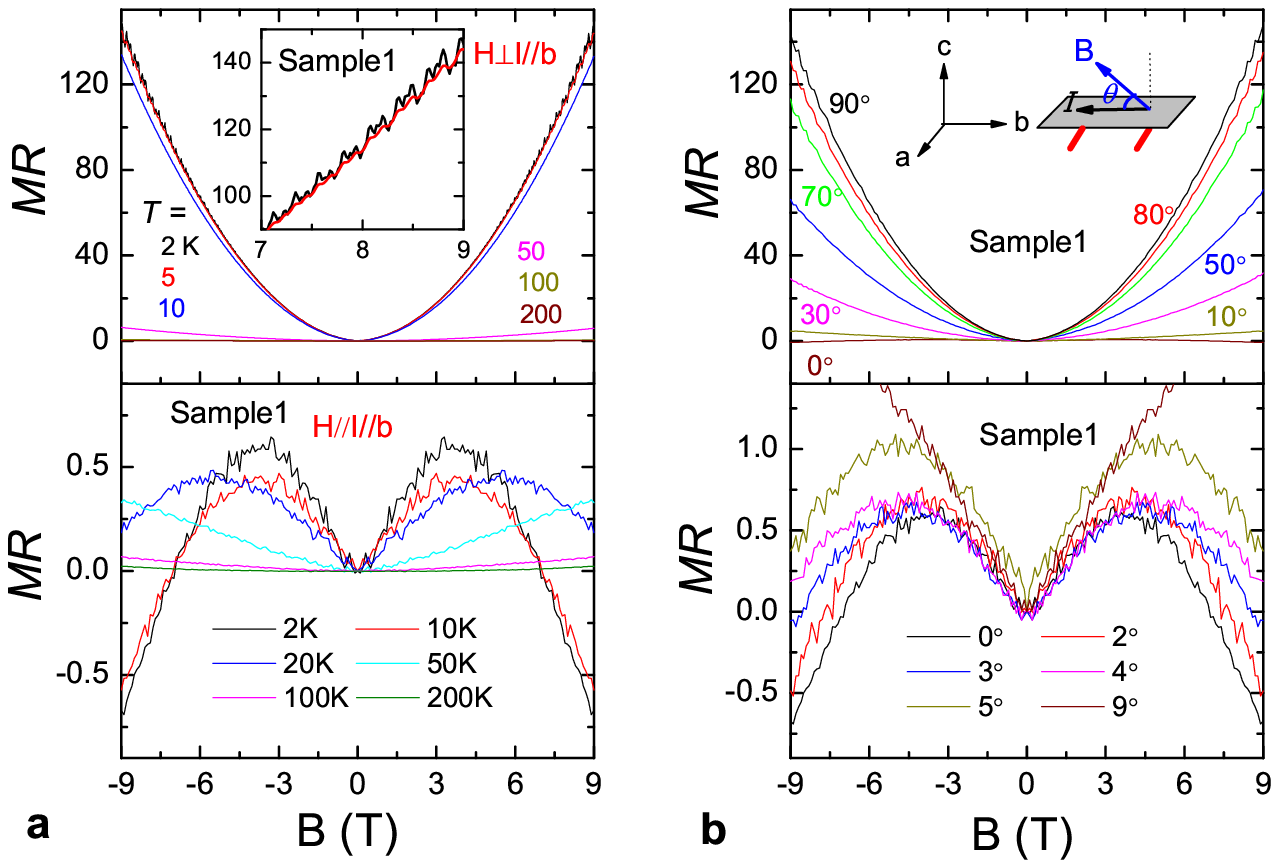}
\caption{{\bf Magnetic field dependence of MR in
TaSb$_2$ single crystal.} {\bf a}, the upper panel:
Magnetoresistance (MR $=(\rho_{xx}(H)-\rho_{xx}(0))/\rho_{xx}(0)\%$) versus
magnetic fields along the c-axis at different temperatures as $
\textbf{B}$ $\perp$ \textbf{I} $\|$ b. The low panel: MR vs. fields for
$\textbf{B} \|\textbf{I} \|$ b. {\bf b}, The upper panel:
MR plotted as a function of magnetic fields at
different angles between $\textbf{B}$ and $\textbf{I}$. The
low panel: the large and unsaturated negative MR emerges in a narrow
window of angle around $\theta = 0^\circ$.}
\end{figure}

\clearpage

\begin{figure}
\includegraphics[angle=0,width=16cm,clip]{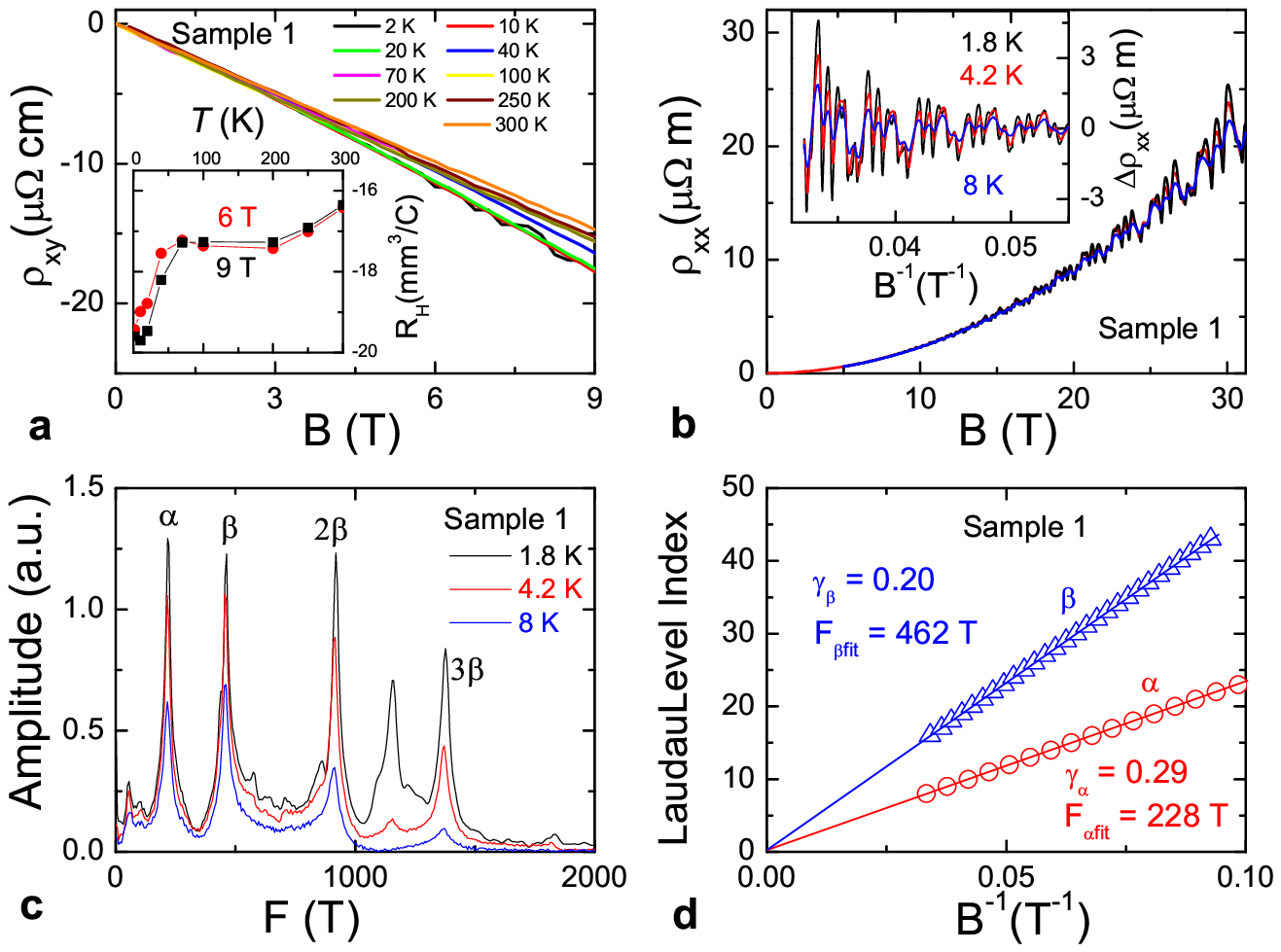}
\caption{{\bf Hall effect and quantum oscillations in sample 1 for
TaSb$_2$.} {\bf a}, Magnetic field dependence of Hall resistivity at
several different temperatures up to 9 T. The inset shows the Hall
coefficient vs. temperatures. {\bf b}, The oscillation vs. high
magnetic field up to 30 T at several temperatures. The inset shows
the oscillatory $\Delta\rho $ vs. inverse field at 1.8, 4.2 and 8 K
within the resistivity plateau regime. {\bf c}, The Fast Fourier Transform
(FFT) spectrum of $\Delta\rho $  at 1.8, 4.2 and 8 K, showing two
oscillation frequencies $F_\alpha =$ 220 T and $F_\beta =$465 T.
{\bf d}, Landau level index plots inverse B versus n. The peak position in $\Delta\rho $ is assigned
as the integer indices.}
\end{figure}


\end{document}